# On the Robustness of Phosphine Signatures in Venus' Clouds

Jane S. Greaves[1], William Bains[2], Janusz J. Petkowski[2], Sara Seager[2], Clara Sousa-Silva[2], Sukrit Ranjan[2], David L. Clements[3], Paul B. Rimmer[4], Helen J. Fraser[5], Steve Mairs[6], Malcolm J. Currie[6].

[1]Cardiff University, Cardiff, UK; [2]Massachusetts Institute of Technology, Cambridge, MA, USA; [3]Imperial College London, London, UK; [4]University of Cambridge, Cambridge, UK; [5]The Open University, Milton Keynes, UK; [6]East Asian Observatory, Hilo, HI, USA.

*Abstract:* We published spectra of phosphine molecules in Venus' clouds, following open-science principles in releasing data and scripts (with community input leading to ALMA re-processing, now benefiting multiple projects). Some misconceptions about de-trending of spectral baselines have also emerged, which we address here. Using the JCMT $PH_3$-discovery data, we show that mathematically-correct polynomial fitting of periodic ripples does *not* lead to "fake lines" (probability < ~1%). We then show that the ripples can be characterised in a non-subjective manner via Fourier transforms. A 20 ppb $PH_3$ feature is ~5σ compared to the JCMT baseline-uncertainty, and is distinctive as a narrow perturber of the periodic ripple pattern. The structure of the FT-derived baseline also shows that polynomial fitting, *if unguided,* can amplify artefacts and so artificially reduce significance of real lines.

*Context*

Our discovery of phosphine in Venus' clouds has sparked much debate – unexpected in an oxidised atmosphere, it suggests a novel chemical origin[1], or even a contribution of life in a hyper-acidic aerial biosphere[2]. Greaves et al.[3] described the detection of the J=1-0 rotational transition of $PH_3$, using both JCMT[i] and ALMA[i] to provide a robustness check; other transitions are difficult to access with ground-based telescopes. Other work has provided $PH_3$ limits[4] and detections[5] at lower and higher altitudes[6].

Several authors[7,8,9] have suggested that the JCMT and/or ALMA data do *not* demonstrate detection of the $PH_3$ 1-0 line, after adapting and re-running our published scripts[3]. These works suggest some misunderstandings of detrending in heterodyne spectroscopy. For example, ref. [7] considered that a 12th-order polynomial fit to ALMA spectra allowed too much freedom – but such a fit is now independently verified in expert follow-up of observatory re-processing[ii,6]. We had already shown[3] that using polynomial fitting did not make $PH_3$ lines "appear", and used a "double-coincidence" test to show that only the $PH_3$ line was in common to both datasets, *without* any polynomial fitting.

However, the original-processing ALMA feature[6] over-estimated $PH_3$, and we infer that a necessary assumption, that ripples do not differ pathologically near the line-frequency, was not met in this dataset. Some of the standard ALMA calibration procedures were recently found to be unsuitable for these observations of an exceptionally bright target[iii]. These issues are now largely mitigated, because after re-calibration, most ripples are wider than the lines of interest; ref. [6] present revised findings.

In Supplementary Information (SI, below), we outline a mathematical approach to polynomial spectral fitting, with examples where applications of our reduction scripts have led to anomalous results[7-9].

*Testing for fake lines*

Refs. [7-9] argue that "fake lines" can be produced by polynomial-fitting, generating populations of features within which they argue the $PH_3$ feature is not significant in amplitude. For clarity: we do not dispute that polynomial fitting can identify instrumental features similar to absorption lines, and indeed amplify them under some constraint-violations (see SI). However, refs [7-9] confuse hypothesis testing with hypothesis generation – essentially assuming that the $PH_3$-identification was a post-hoc rationalisation of a feature found after complex processing. In fact, Venusian-$PH_3$ was a prior hypothesis that we tested with data[iv]. Specifically, the line's position was predicted, and so any "fake lines" had not only to show an absorption but also show an absorption at a pre-specified location.

---

[i] JCMT: James Clerk Maxwell Telescope; ALMA: Atacama Large Millimeter/submillimeter Array.
[ii] In this experimental bandpass-self calibration of the Venus visibility data, 12th and 5th-order polynomials in amplitude and phase respectively were found to be optimum; in ref. [3] we post-processed extracted spectra.
[iii] See https://www.eso.org/public/unitedkingdom/announcements/ann20030/.

Our tests[3] were: (a) a planetary absorption line should lie where expected from the transition's rest-frequency *and* the Doppler-shift, (b) the line-width should be appropriate to the atmospheric physics, and (c) negative amplitude versus the continuum is required. These tests are not all likely to be met by a random ripple, although the probability does increase if there are many ripples in the passband.

Here (a) is the most important criterion, with (b) less rigorous: the $PH_3$-$CO_2$ collisional broadening coefficient is unmeasured, so the line-width is loosely constrained. We here test (a) by examining how many "draws" are needed to yield a feature coincident with pre-specified locations in the JCMT spectrum. Any random test-frequencies could be used, but we focussed on rest-frequencies of molecules that are improbable in Venus' clouds (C-bearing compounds, see SI). This can additionally show whether potential mis-identifications could occur, e.g. only one of a family of lines is seen.

Figure 1 shows that very few draws produced "absorptions" that agreed with the predicted location. There are only three candidates, with two mapping onto the largest instrumental ripples we have characterised in the passband (see next section), and the third is an unconvincing fit (with 50% uncertainty in the derived width). We thus estimate that the chance of any instrumental feature passing our tests, and not being readily recognised as an artefact, is under ~1 % (<1 among 95 draws).

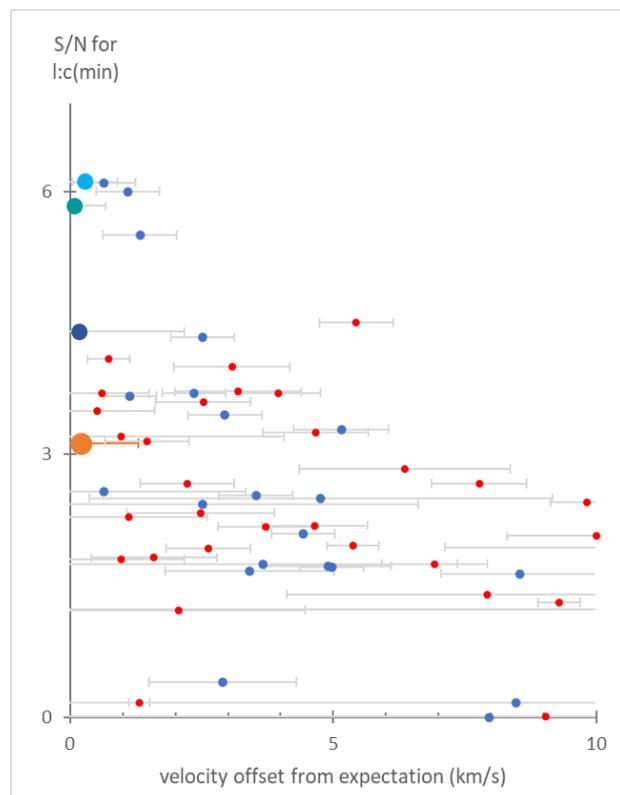

***Figure 1.*** *(Right:) S/N, signal-to-noise ratio at line:continuum minimum, plotted against the (absolute) offset of the line centre from the velocity a transition would be expected to have in the Venus reference frame. Red and blue symbols are respectively for draws giving a positive result ("emission") or a negative result ("absorption"). The orange point is $PH_3$ 1-0 from Table 1 of ref. [3]; other points were auto-characterised using Lorentzian line-fits. Only the three points shown with larger symbols agree within errors with the expected velocity and are in absorption, with S/N > 3. (Below:) For these three extracted features, the central parts of the spectra (histograms) and their Lorentzian fits (dashed) are plotted over the net baseline (black curve) defined by Fourier components. The thicker black sections highlight where two features (light blue, turquoise) coincide with the largest ripples; the third feature (dark blue) is poorly defined (uncertainty of ±50% in width).*

---

[iv] Abstract at almascience.eso.org/observing/highest-priority-projects, under 2018.A.00023.S (Cycle 6 DDTs).

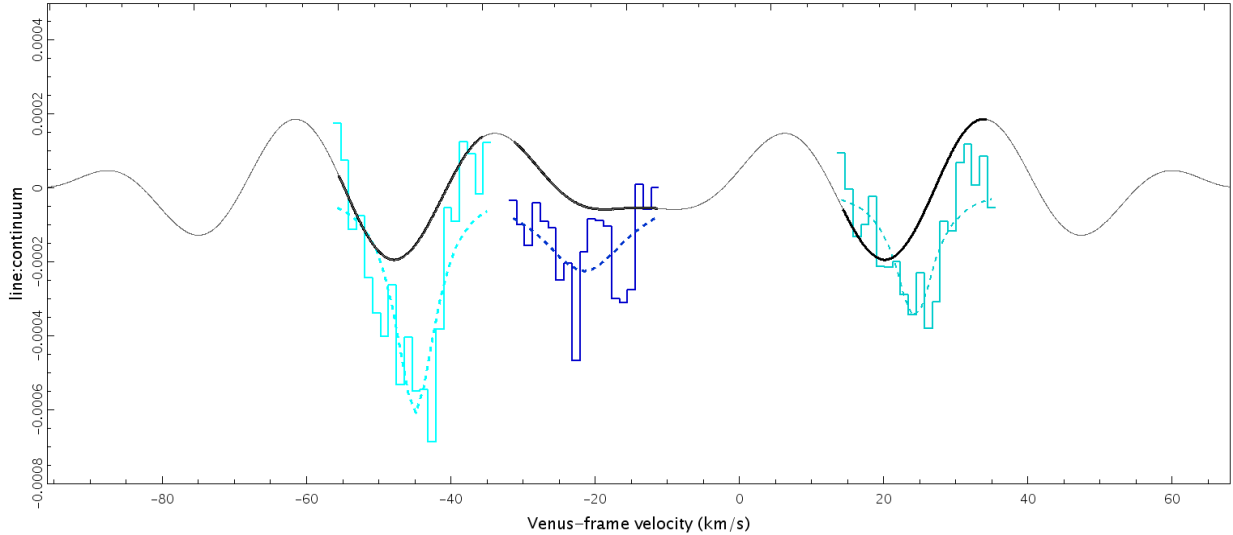

Ref. [9] performed JCMT-draws using the same scripts, and concluded that 25% of them could produce a feature of larger amplitude than the PH$_3$ line. Here we find 20% of draws in absorption and deeper than PH$_3$, but only ~2% (2/95) have as good positional-agreement and velocity-uncertainty as the real PH$_3$ line, and these are clearly instrumental. Also, higher-amplitude cases in ref. [9] likely arise where the ripple is largest, and using a constant polynomial order gives excess freedom (see next section).

*Fourier analysis of baselines*

Ref. [7] argued that the ALMA PH$_3$ line was non-significant, at ~2σ level, after flattening the (original) ALMA spectra with low-order polynomials, and treating residual systematics as a random noise term. This ALMA processing is now superseded, but we use the JCMT data here to show that a feature that is "low-σ" by ref. (7)'s definition can in fact be a significant outlier among non-random *periodic* ripples.

As described in the SI, Fourier transform (FT) components can be identified in the frequency-time plane of the JCMT dataset. We identify six major components, >10 standard-deviations above the FT-noise. These arise from four different ripple periods and two periodicities over time. An inverse-FT of these components then generates a set of spectral baselines, with *no* subjective intervention.

In Figure 2, we show the net baseline from this procedure, along with the dispersions that result because the ripples are not the same in each observation. The superposed PH$_3$ model (20 ppb) is then a clear outlier – it perturbs the periodicity of the ripples, and is narrower than the ripple "bumps".

Here, the line-minimum would be only 3σ by the definition of ref. [7], i.e. comparing it to the standard deviation of the ripple pattern. However, it is 4.8σ compared to uncertainty in the spectral baseline, allowing for a significant detection if the measurement noise is small. In practice, the JCMT spectrum[3] had substantial per-channel noise, but up to ~7σ was achieved by integrating over the line.

There are some sections of model-baseline (Figure 2) where the ripple pattern is flat within the noise, e.g. around ±(70-90) km/s. This can explain some of the results of ref. [9], where some polynomial fits yielded features of higher amplitude than the PH$_3$ line. The constant order provided in our scripts (8[th], optimised for the PH$_3$ line-region) will allow too much freedom in a section of passband with a lower number of distinguishable bumps. In such cases, artificially amplified features can easily occur.

This Fourier analysis intentionally uses *minimal* intervention, and is intended only to demonstrate an approach that is non-subjective. In reality, the ripple artefacts changed between observations, in number of periods seen and in their amplitudes and phases, so line-extraction needs to be done per-observation. The number of >5σ FT-components in individual observations is 2-8, so the more complex cases approach the number of parameters (9) used in the polynomial fits[3]. However, further-developed FT-solutions may be able to both remove subjectivity and flatten the entire passband.

*Figure 2. Fourier-component model for the net JCMT baseline (in red), with 20 ppb PH$_3$ model added (in black). The PH$_3$ radiative-transfer model was shifted to the instrumental velocity frame, and median-filtered over 35 km/s intervals as in ref. [3]. Red bars show the standard errors derived from the dispersion among the model baselines of different observations. The line-minimum signal is -2.65 10$^{-4}$ and the baseline uncertainty in this spectral channel is 0.55 10$^{-4}$. The dashed blue box indicates an example of a window in the passband that has a low number of distinguishable ripples: the three bumps around -70 to -90 km/s merge into one feature of the same signal-level within the uncertainties.*

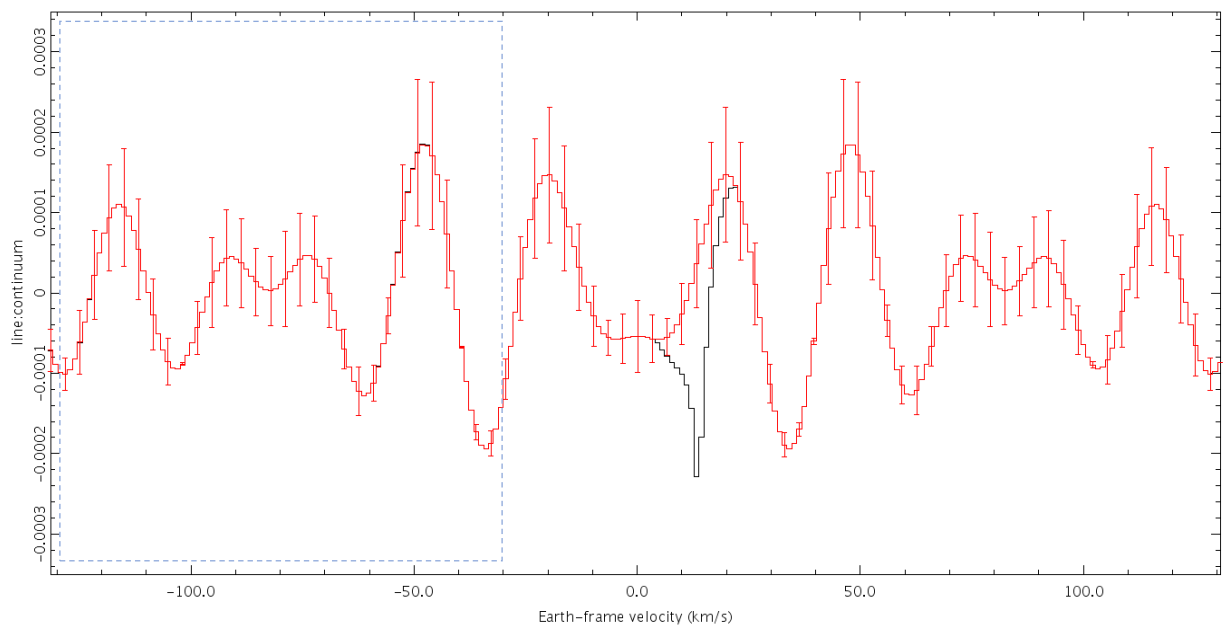

## Supplementary Information

*Mathematical approach to fitting spectral baselines*

Here we clarify two points about fitting polynomials to bandpasses with instrumental ripples, following standard ideas in millimetre-waveband heterodyne spectroscopy.

1. The appropriate polynomial order is P = 1+N, for N peaks-plus-troughs observed in a passband, and excluding the interpolation region within which the line resides. For example, $y=x^2$ is the minimum order that can fit one instrumental "bump". A corollary is that when interpolating across a fraction Z of the passband, the number of peaks-plus-troughs present is reduced to ≈(1-Z)N, if the ripples are periodic. If P is > [1 + (1-Z)N], the polynomial has excess freedom and so "fake lines" are readily generated. If P is < [1 + (1-Z)N], the spectral residuals will be larger than necessary, which will decrease the significance of the line compared to residuals.
2. For correct interpolation across the section where the line exists, there needs to be some information on whether signals are trending up or down immediately adjacent to the line. This is essentially a Nyquist-sampling argument: sampling-points need to be spaced in frequency by 2B in order to characterise a ripple of period B. In the JCMT data, our preferred fit[3] interpolated over 10 km/s, approximately at the Nyquist limit for ripple periods ~18 km/s.

In the analyses by other authors[7-9], discrepancies appear. Ref. [7] applied a 12th-order polynomial to the original ALMA processing, but half of the examples (their Figure 2) have no data across a substantial section (in offsetting their bands, they include frequencies where nothing was recorded). This contravenes (1), i.e. P is > [1 +(1-Z)N], making spurious features likely. Similarly, ref. [8] showed (their Figure FS1) application of a 12th-order polynomial to the original-processing *wideband* ALMA data – which are dominated by systematic ripple, prohibiting any line-extraction[6]. Their use of P = 12 also contravenes (1) as only ~7 peaks and troughs are present, and although (2) appears not formally contravened, their fit does in fact generate a feature of width ~B. Finally, ref. (9) explored some P = 3-4 fits on JCMT-spectrum sections totalling 50 km/s (their Figure 2), while ~6 peaks-plus-troughs are expected in this velocity-range for ~18 km/s-period ripples. Hence these tests contravene (1) by using P that is < [1 + (1-Z)N], and so will leave residuals that artificially reduce the real line's significance.

*Procedure for sampling for "fake lines"*

Our JCMT passband[3] was 250 MHz, corresponding to 281 km/s at the $PH_3$ 1-0 frequency of 266,944.5 MHz. After rejecting 256 noisier channels at each end of the full passband of 8192 channels, and allowing ±(5-50) km/s fitting-regions around each target-frequency, 153 km/s of passband remains for extraction tests. The accuracy in position of the $PH_3$ line in ref. [3] was ±1.1 km/s, so the number of fully-separable test-regions is 153/2.2 = 70. In ref. [9], extractions were spaced by the correlator resolution of 0.03427 km/s, so many of these draws yielded similar results.

Here we generated a set of test-frequencies by searching the JPL online spectroscopy catalogue[v]. A random-number generator could be used, but adopting real line frequencies also allows for testing of confirmation bias, if multiple similar transitions exist in the passband. (This "missing features" test is classic in the field, e.g. in the search for interstellar glycine.) We then selected all C-bearing species in the usable band, since organics are unlikely in Venus' atmosphere. The resulting catalogue is intrinsically of random molecular species; it finally included 25 molecules (and isotopic-substitutes) of 4 to 13 atoms, with HClCO being the smallest and $C_3H_8O_2$ the largest.

In the line-list, we have removed frequencies uncertain by > 1 MHz (1.1 km/s), or with closest-neighbour transitions within 1 MHz, or where the interpolation region[3] of ±5 km/s would overlap with that of the $PH_3$ line. As only 56 transitions then remained, frequencies were "cloned" via increments of 1 MHz where this could retain a gap of 1 MHz to the next nearest catalogue entry. This added 39 frequencies, so the final test-set was of 95 spectral extractions ("draws"). Our script was then run over windows of 100 km/s centred around the target-frequencies, and extracted features were fitted with Lorentzian functions (over ±10 km/s) to estimate line-minimum, line-centre, and full-width at half-minimum (FWHM). Figure 1 plots the fitted line-minimum against the offset of the fitted line-centre, i.e. its offset from the transition-frequency when Doppler-shifted to the Venusian reference frame. The two features with best positional coincidence (light blue, turquoise points) are in fact test-molecules where only 1 of 2 possible transitions was seen, thus failing a "missing-features" test.

*Fourier analysis of JCMT spectra*

The observations with receiver RxA3 on JCMT were described in detail in ref. [3]. The periodic sinusoidal ripples are mainly related to the telescope and its surroundings, rather than the receiver. During our observations in the period 9-16 June 2017, the spectral ripples were seen to be unstable, changing both during and between observing-mornings. Changes in phase, amplitude, and number of components are apparent, possibly due to temperature variations or flexing of cables.

---

[v] See https://spec.jpl.nasa.gov/cgi-bin/catform.

The Fourier transform used to make Figures 1 and 2 was performed using the KAPPA[10] task *fourier*, on a stack of all 140 sub-observations ordered contiguously in time[vi]. An FT-Hermitian frame was made with pixel-values describing phase plus amplitude; all pixels in the Hermitian except for those describing the major ripples were then downweighted (by 100,000), and an inverse-FT was made to generate a set of 140 spectral baselines. The retained Hermitian components have absolute pixel-values of 4.67-6.39, above 10σ for a standard deviation of 0.42; see Figure SI-1. Collapsing the stack of baselines down the time axis gave the net-baseline model. Uncertainties were generated using the KAPPA task *setvar*, and describe the dispersion of the baselines in the input observations.

*Figure SI-1. Top panel: Time-versus-frequency stack of the JCMT data, after removal of reflection-artefacts: footnote(vi). Frequency channels (x-axis) have been binned in groups of 32 for clarity. Each row represents a 2-minute data-sample; observations comprised 14 samples. Middle panel: FT Hermitian of the upper panel data; pixels show components around (0,0) out to periods of 32, 10 in the frequency, time axes respectively. Lower panel: histogram of Hermitian values. The six points shown in red are the >10σ Hermitian-outliers used to generate the model spectral baseline in Figures 1, 2.*

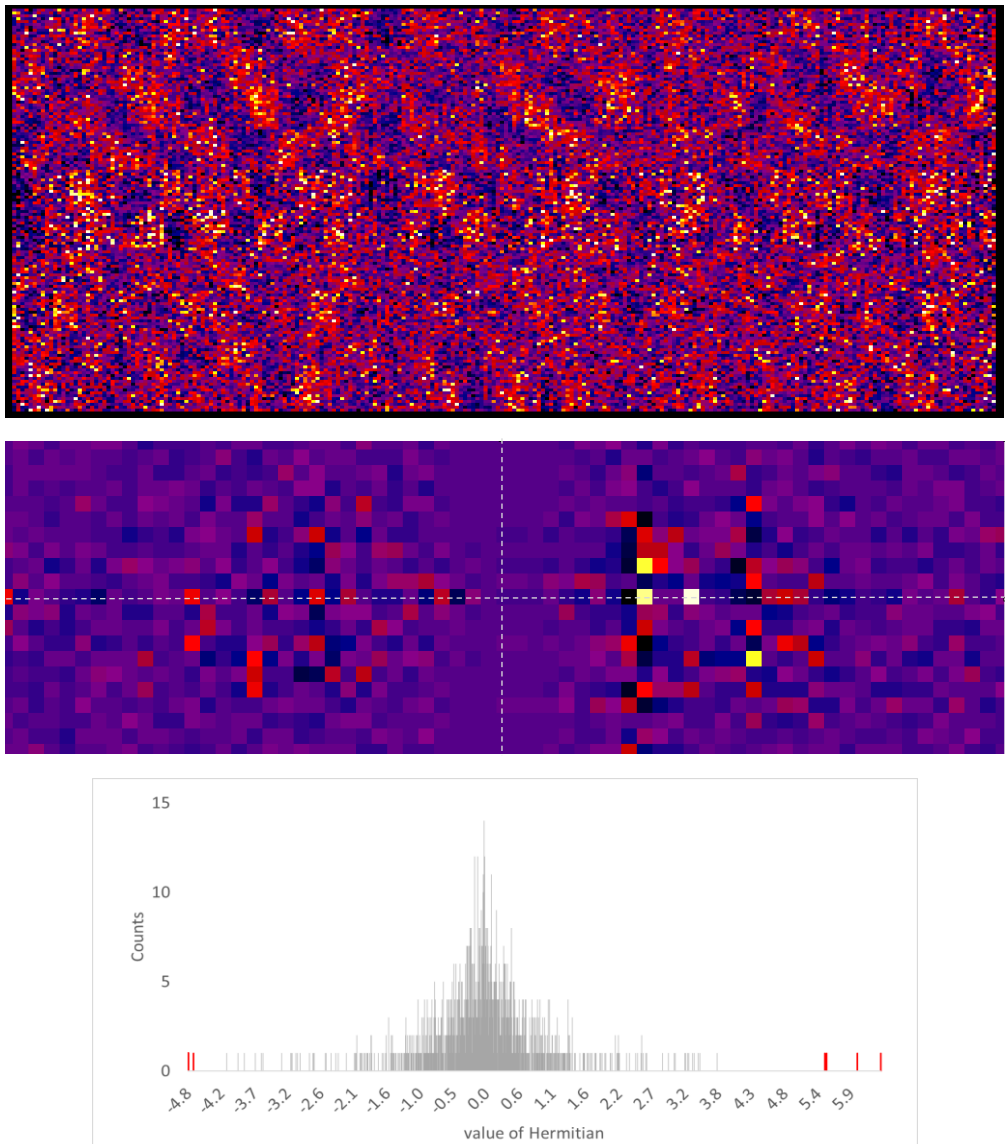

---

[vi] This procedure was implemented after the removal (via polynomials) of broad waves due to reflections, which have periods wider than ~60 km/s, as described in the SI of ref. [3].

The principal waves found by this method have 8, 9, 12 and 16 periods across 250 MHz of passband. The shortest-period wave is the known "16 MHz ripple", and other components may be modulations of this effect. Two effects over time were also identified, with the 16-period ripple changing on inter-observation timescales, while the 9-period ripple varied between observing mornings.

Ripple periods have also been identified for individual observations, by noting that the real part of the FT is maximised when a number of sinusoids *exactly* fit in the passband. By systematically cropping channels from the passband ends, real-FT peaks could be identified where the pixels describing ripple-amplitudes exceeded ~3σ. The mean periods found in the JCMT data were 29.5, 20.5 and 15.8 MHz. Not all components are present in all observations, and periods varied by up to ~7% from the means.